\documentstyle[12pt]{article}
\begin{document}
\baselineskip 19pt plus 1.5pt
\hspace{9.5cm}
BRX-TH-345 \\
           \vspace{0.4cm}
           \begin{center}
           \begin{Large}
           \begin{bf}
4D and 2D Evaporating Dilatonic \\
Black Holes \\
            \end{bf}
            \end{Large}
\vspace{0.8cm}
Yoav Peleg\footnote{This work is supported by a Fishbach Fellowship
and by the NSF grant PHY 88-04561.} \\
\vspace{0.4cm}
Physics Department, Brandeis University, Waltham, MA 02254-9110\\
\vspace{0.2cm}
{\em Peleg@Brandeis}\\
\vspace{1cm}
           \begin{large}
ABSTRACT\\
            \end{large}
            \end{center}

The picture of S-wave scattering from a 4D extremal dilatonic
black hole is examined. Classically, a small matter shock wave
will form a non-extremal black hole. In the ``throat region"
the $r-t$ geometry
is exactly that of a collapsing 2D black hole. The 4D Hawking
radiation (in this classical background) gives the 2D Hawking
radiation exactly in the throat region.
Inclusion of the back-reaction changes
this picture: the 4D solution can then be
matched to the 2D one only if the Hawking radiation is very small
and only at the beginning of the radiation. We give (explicitly) that
4D solution. When the total radiating energy approaches the energy carried
by the shock wave, the 4D picture breaks down. This happens even
before an apparent horizon is formed, which suggests that the 4D
semi-classical solution is quite different from the 2D one.

\newpage

\section{Introduction}
There is much current interest in both the classical and
quantum aspects of dilatonic black holes in various dimensions.
Let us review some relevant ideas. Consider
the extremal 4D magnetically charged black hole which has two asymptotic
spatial regions. One is far away from the ``black hole", and the
other is ``down the infinite throat" [1]. The 4D manifold down
the infinite throat is $ \Sigma^{4} = \Sigma^{2}
\times S^{2} $, where $ S^{2} $ is a 2-sphere of constant radius.
By a Kaluza-Klein process one can get from it an effective 2D
theory, defined on $\Sigma^{2}$ [2].
A general static solution of this 2D theory is a black hole [2,3].
Including matter fields, one can describe a collapse process, and
semiclassical Hawking radiation [4]. The backreaction
calculations can be handled in 2D. By extending the
model [5-7], one can even solve exactly the one loop backreaction equations.
The solutions describe formation and evaporation of a black hole.
But there are still some open questions concerning the ``end point" of
the process [8,9] and the mass definition [10-13], so the information
puzzle is still unresolved.

Is there a 4D interpretation of those 2D results? The standard
picture is that the 2D process is the $r-t$ part of a 4D
S-wave scattering. In this paper we will examine this picture.
It is very easy to find the relation between the 4D and 2D
static black holes, and even to describe a classical
collapsing star.
Hawking radiation (in the classical background of a
collapsing star) is well known in both 4D [14] and 2D [4], the
relation between them being consistent with the S-wave picture,
as we will see. When the back-reaction is taken into
account, a simple S-wave scattering is consistent
only with very small Hawking radiation, and only at the
beginning of the radiation process, before the radiation carries
away the total energy of the matter shock wave. In that case one
can give an explicit 4D interpretation. Namely, one can find a 4D
solution (including the back-reaction) that will correspond to
the 2D evaporating star. If the Hawking radiation
is not very small, it is not clear whether there is a 4D
interpretation. It is not clear if one can find a 4D
solution (including the backreaction) that corresponds
to the 2D evaporating star. In that case the 4D
``1 loop" Einstein equations,
$ G_{\mu \nu}^{(4D)} = <T_{\mu \nu}^{(4D)}> $ , are
quite complicated, unlike the case of very small Hawking
radiation, in which those equations become very simple.

The structure of the paper is as follows:
In section 2, we describe the relation between the 4D and 2D
static dilatonic black holes, and we will see the explicit relation
between the almost extremal 4D black hole and the 2D one. In
section 3, we will describe the classical shock wave collapsing
star (in 4D and 2D). In section 4, the Hawking radiation (in
the classical background) is considered. In section 5, we will
study the backreaction problem. Concluding remarks are in section 6.

\section{4D and 2D Static Dilatonic Black Holes}
The 4D action describing dilaton gravity is [1]
   \begin{equation}
S^{(4)} = \frac{1}{2\pi} \int d^{4}x \sqrt{-g} e^{-2 \phi} \left(
R^{(4)} + 4 (\nabla \phi)^{2} - \frac{1}{2} F^{2} \right)
   \end{equation}
The magnetically charge spherical symmetric static solution is
   \begin{eqnarray}
ds^{2} &=& e^{2\phi_{0}} \left( -\frac{(1 - r_{_{H}}/r)}{(1 - r_{s}/r)}
dt^{2} + \frac{dr^{2}}{(1 - r_{_{H}}/r)(1 - r_{s}/r)} + r^{2} d\Omega_{2}^{2}
\right) \nonumber\\
F_{\theta \phi} &=& Qsin\theta \nonumber\\
e^{-2\phi} &=& e^{-2 \phi_{0}} (1 - r_{s}/r) ~.
   \end{eqnarray}
Here $M$ and $Q$ are the mass and charge of the black hole, $ r_{_{H}} $
is the horizon, $ r_{_{H}} = 2M $ , and $ r_{s} $ is the singularity,
$ r_{s} = Q^{2}e^{2\phi_{0}}/M $ . To prevent a naked singularity,
we require $ r_{_{H}} \geq r_{s} $ (or $ 2M^{2} \geq Q^{2}e^{2\phi_{0}} $,
the equality defines the extremal black hole).
There are three regions in this space-time: 1) the exterior,
$ r > r_{_{H}} $. 2) $ r_{_{H}} > r > r_{s} $, which is inside the
black hole. 3) $ r_{s} > r > 0 $; this is an "extra region", which
includes a naked singularity. Region 3) is of course disconnected (and
``irrelevant" to observations made in 1) and 2).).
We will be interested only in regions 1) and 2), which describe the black
hole, so the radial coordinate
   \begin{equation}
\tilde{r} \equiv r - r_{s}
   \end{equation}
will be non-negative. It is convenient to define
the deviation from the extremal solution
   \begin{equation}
\epsilon \equiv \frac{r_{_{H}}}{r_{s}} - 1
   \end{equation}
In terms of these, the metric becomes
   \begin{equation}
ds^{2} = e^{2\phi_{0}}\left( - (1 - \epsilon r_{s} / \tilde{r} )
dt^{2} + \frac{{(\tilde{r}/r_{s} + 1)}^{2} d\tilde{r}^{2}}{(1-
\epsilon r_{s} / \tilde{r} ){(\tilde{r}/r_{s})}^{2} } + {(\tilde{r} +
r_{s} )}^{2} d\Omega_{2}^{2} \right)
   \end{equation}
and the dilaton field is
   \begin{equation}
e^{-2\phi} = e^{-2\phi_{0}}\frac{\tilde{r}}{(\tilde{r} + r_{s})}~.
   \end{equation}
We can see from (5) and (6) that for $ \tilde{r} << r_{s} $
   \begin{eqnarray}
ds^{2} &\longrightarrow & e^{2\phi_{0}}\left(
- (1 - \epsilon r_{s} / \tilde{r} )
dt^{2} + \frac{d\tilde{r}^{2}}{ (1-
\epsilon r_{s} / \tilde{r} ) {(\tilde{r}/r_{s})}^{2}} +
r_{s}^{2} d\Omega_{2}^{2} \right) \\
e^{-2\phi} &\longrightarrow & e^{-2\phi_{0}} \tilde{r}/r_{s}~.
   \end{eqnarray}
Thus, near $r_{s}$, the 4D manifold reduces to
$ \Sigma^{4} = \Sigma^{2} \times S^{2} $, where $S^{2}$
is a 2-sphere, with constant radius $R=e^{\phi_{0}}r_{s}$.
The $ \Sigma^{2} $ part represents a 2D black hole as long as
$ \epsilon << 1 $. One can see that directly from (7), but
we will see this explicitly in the following.

In the region $ \tilde{r} << r_{s} $ we can use the standard
Kaluza-Klein procedure to get (from (1)) the 2D action (on
$\Sigma^{2}$)
   \begin{equation}
S^{(2)} = \frac{1}{2\pi}\int d^{2}x \sqrt{-g} e^{-2\phi} \left(
R^{(2)} +  4 {(\nabla \phi )}^{2} + 4 \lambda^{2} \right)
   \end{equation}
where
   \begin{equation}
4 \lambda^{2} = \frac{2}{R^{2}} - \frac{1}{2}F^{2}(r_{s}) =
\frac{e^{-2\phi_{0}}}{r_{s}^{2}} - \frac{Q^{2}}{r_{s}^{4}}
   \end{equation}
There is also a vector field term (from the Kaluza-Klein
reduction) that we did not write explicitly in (9),
because we will not consider excitations in it.
Using conformal gauge, $ g_{++}=g_{--}=0$, $g_{+-}= -\frac{1}{2}
e^{2\rho} $, the static black hole solution (sometimes called
Witten's b.h.) of (9) is [3,4]
   \begin{equation}
e^{-2\phi} = e^{-2\rho} = \frac{m}{\lambda} - \lambda^{2}x^{+}x^{-}
   \end{equation}
where $ x^{+} , x^{-} $ are the light-cone coordinates for which
the metric is $ g_{+-}=-\frac{1}{2}e^{2\rho}$, and $m$ is the 2D
ADM mass\footnote{One should not confuse the 2D mass, $m$,
with the 4D mass, $M$. The relation between them will be given later.}.
In the coordinates $(t,\hat{r})$
   \begin{eqnarray}
t &\equiv &  \frac{1}{2} ln(-x^{+}/x^{-}) \\
2 \lambda^{2} \hat{r} &\equiv &  e^{-2\phi(x^{+},x^{-})}
   \end{eqnarray}
the 2D metric is
   \begin{equation}
ds^{2} = \lambda^{-2} \left( -(1 - m/2
\lambda^{3} \hat{r} )dt^{2} +  \frac{d\hat{r}^{2}}{
(1 - m/2\lambda^{3}\hat{r} ) {(2\hat{r})}^{2} }
\right)
   \end{equation}
We see that the $ \Sigma^{2} $ part of the 4D solution (7),(8) and the
2D solution (13),(14) have the same form, but with different
parameters. Before comparing them, we must remember that
the 4D solution (2) is a three parameter solution
$ ( M , Q , \phi_{0} ) $ , while the 2D solution (11) is a
two parameter solution $ ( m , \lambda ) $ .
The reason for that is that $r_{s}$ (the radius of the 2-sphere),
which is a function of the three 4D parameters, is a scale that
cannot appear in the 2D solution.
We therefore should fix this scale before relating the solutions.
After choosing the coordinates (12),(13) it is convenient to take
   \begin{equation}
r_{s} = \frac{Q^{2}e^{2\phi_{0}}}{M} = \frac{1}{2}~.
   \end{equation}
When this scale is fixed, we get from (10) and (15) that
$ \lambda^{2} = e^{-2\phi_{0}} (1 - \epsilon) $, and
the 4D solution (7),(8) become the 2-parameter set
   \begin{eqnarray}
ds^{2} & \longrightarrow & \frac{(1-\epsilon)}{\lambda^{2}}
\left( - (1 - \epsilon / 2 \tilde{r} ) dt^{2} +
\frac{d\tilde{r}^{2}}{(1 - \epsilon / 2 \tilde{r})
{(2\tilde{r})}^{2} } + \frac{1}{4} d\Omega_{2}^{2} \right) \\
e^{-2\phi} & \longrightarrow & \frac{2 \lambda^{2} \tilde{r}}{1 - \epsilon}
   \end{eqnarray}
We can see from (13),(14) and (16),(17) that if we identify
   \begin{equation}
\hat{r}=\tilde{r} ~~~,~~~\epsilon = \frac{m}{\lambda^{3}}
   \end{equation}
then for $ \epsilon << 1 $ , $\Sigma^{2}$ is exactly the 2D Witten
black hole.

At this point we have explicit relations describing the well known
picture: the 2D black hole solution is the $r-t$ part of a
4D {\em almost} extremal black hole ($ 0 < \epsilon << 1 $) in the region
``down the throat" ($ \tilde{r} << r_{s} $).
The 2D mass, $m$, is really the mass deviation from the 4D extremal black
hole (18). As was noticed by Witten [3],
the 2D zero mass solution (the linear dilaton solution) represents
an extremal 4D black hole and not a flat 4D space-time.

Outside the region $ \tilde{r} << r_{s} $,
the 4D and 2D solutions are quite different. For example $\tilde{r} $
but not $\hat{r}$, is an asymptotically flat coordinate (see (5)). So
if we want to give a 4D interpretation to the 2D results, we
must restrict ourselves to $ \tilde{r} << r_{s} = 1/2 $
(or $ \hat{r} << 1 $). Using (11) and (13) we see that this
means that $ x^{+}x^{-} << 1 $. Let $ \tilde{r}_{c} = \hat{r}_{c}
<< 1$ be the radius at which we ``glue" the 4D solutions in the
following sense For
$ \tilde{r} \leq \tilde{r}_{c} $ we will use the 2D results to
describe the $r-t$ part of the 4D space-time. But we cannot do that for
$ \tilde{r} \geq \tilde{r}_{c} $ ; in that region one must solve
the 4D equations. In the ($x^{+},x^{-}$) plane, $\hat{r}=\hat{r}_{c}$
is a line $x^{+}x^{-} = const. << 1$.

The proper distance from $\tilde{r} =
\tilde{r}_{c} $ to the singularity ($\tilde{r} = 0 $) goes like
$ ln(1 + c\tilde{r}_{c}/\epsilon ) $ where $c$ is some constant.
So if we want a long throat, $ \tilde{r}_{c}/\epsilon $
must be much bigger then 1, which means $ \epsilon << \tilde{r}_{c}
<< 1 $. So from now on we can neglect $\epsilon$
( but not $ \epsilon/\tilde{r}_{c} $ or $\tilde{r}_{c} $) relative to $1$.

\section{Classical Collapsing Star}
Adding matter fields to (1) or (9) enables us to find
solutions that describe a collapse process. A simple collapse
process can be described by an ``$f$-shock wave" (The $f$ fields
are zero everywhere but at $ x^{+}=x^{+}_{0} $). The solution is
the extremal black hole ( $ m = 0 $ (or $\epsilon = 0$ ))
for $ x^{+} < x^{+}_{0} $, and non-extremal
black hole ($ m > 0 $) for $ x^{+} > x^{+}_{0} $. The
energy carried by the $f$ fields is $m$.

First consider the 2D case [4]. The 2D classical action
acquires a kinetic term for the matter fields $f_{i}$,
   \begin{equation}
S^{(2)} = \frac{1}{2\pi}\int d^{2}x \sqrt{-g} \left[
e^{-2\phi} \left(
R^{(2)} +  4 {(\nabla \phi )}^{2} + 4 \lambda^{2} \right)
- \frac{1}{2} \sum_{i=1}^{N} {(\nabla f_{i})}^{2} \right]
   \end{equation}
The classical collapsing solution is
   \begin{eqnarray}
& & f_{i} = 0 ~~~,~~~ x^{+} \neq x^{+}_{0} \nonumber \\
& & e^{-2\phi} = e^{-2\rho} = -\lambda x^{+}x^{-} -
    \frac{m}{\lambda x^{+}_{0}}( x^{+} - x^{+}_{0} ) \Theta
    ( x^{+} - x^{+}_{0} )
   \end{eqnarray}
We see that this solution describes a linear dilaton solution
for $x^{+}<x^{+}_{0}$, and a 2D black hole (with mass $m$)
for $x^{+}>x^{+}_{0}$.
The classical energy momentum tensor of the $f$ fields is
$ T^{(f)}_{++} = \frac{m}{\lambda x^{+}_{0}} \delta (x^{+} -
x^{+}_{0} ) $ (an incoming shock wave).

Is there a corresponding 4D solution? It should be the extremal black
hole ($\epsilon = 0$) for $x^{+}<x^{+}_{0}$, and non-extremal
black hole ($\epsilon > 0$) for $x^{+}>x^{+}_{0}$. It is
very easy to see that this is indeed the case. For $\tilde{r}
> \tilde{r}_{c}$ one must solve the 4D equations of motion,
implied by (1) (with the matter fields)
   \begin{eqnarray}
& & e^{-2\phi} ( R_{\mu \nu} + 2 \nabla_{\mu} \nabla_{\nu}
\phi - F_{\mu \delta}{F^{\delta}}_{\nu} ) +
\pi  T^{C}_{\mu \nu} = 0 \\
& & 4 \nabla^{2} \phi - 4 {(\nabla \phi )}^{2} + R - \frac{1}{2}
F^{2} = 0  \\
& & \nabla_{\mu}( e^{-2\phi} F^{\mu \nu} ) = 0
   \end{eqnarray}
where $ T^{C}_{\mu \nu} $ is the classical matter energy momentum
tensor.

In our case $ \frac{\epsilon}{\tilde{r}_{c}} << 1 $, and we see
from (5) that for $ \tilde{r} \geq \tilde{r}_{c} $ the metric
changes only slightly relative to the extremal metric ($\epsilon
= 0$ ). So we can use the {\em linearised} equations.
The background ``vacuum" metric is the extremal
black hole whose line element (see (5)) is
   \begin{eqnarray}
d s_{(0)}^{2} &=& g_{\mu \nu}^{(0)} dx^{\mu} dx^{\nu} =
e^{2\phi_{0}} \left( - dt^{2} +
{ \left( 1 + \frac{r_{s}}{\tilde{r}} \right)}^{2} d \tilde{r}^{2}
+ {(\tilde{r} + r_{s} )}^{2} d \Omega_{2}^{2} \right) \nonumber \\
&=&  e^{2\phi_{0}} \left( - dt^{2} +  {dr^{*}}^{2}
+ {(\tilde{r}(r^{*}) + r_{s} )}^{2} d \Omega_{2}^{2} \right)
   \end{eqnarray}
where $ r^{*} = \tilde{r} + r_{s} ln(\tilde{r}) $. The linear
deviations from the ``vacuum" metric, dilaton and EM fields, are
defined by
   \begin{eqnarray}
g_{\mu \nu} &=& g^{(0)}_{\mu \nu} + h_{\mu \nu} \nonumber \\
h_{\mu \nu} &=& e^{2\phi_{0}} diag ( \delta , \sigma ,
{(\tilde{r} +r_{s})}^{2}\eta , ({(\tilde{r} + r_{s})}^{2} sin^{2}\theta)
\eta ) \nonumber \\
\phi(t,\tilde{r}) &=& \phi_{0} -\frac{1}{2} ln\left(
\frac{\tilde{r}}{\tilde{r}+ r_{s}} \right)
+ \gamma(t,\tilde{r}) \nonumber \\
F_{\theta \phi} &=& Q sin\theta (1 + \rho)
   \end{eqnarray}
where $ \delta ~,~ \sigma ~,~ \eta ~,~ \gamma $ and $ \rho $, are
all much smaller then one.
The linearised form of (23) using (25) leads to $ \eta \simeq 0$,
where in our approximation $ \simeq 0 $ means of order $\epsilon$
or $ {(\epsilon / \tilde{r}_{c})}^{2} $.
Equation (22) is just the Bianchi identity obeyed by (21), so
one should not consider it as an independent equation. The
non-vanishing components of the linearised Einstein equations (21),
are
   \begin{eqnarray}
(tt) ~ & & ~ \frac{1}{2}( \ddot{\sigma} + \delta'' )
+ \frac{(\tilde{r}
+ r_{s}/2)}{{(\tilde{r} + r_{s})}^{2}} \delta' -
2 \ddot{\gamma} =
\pi e^{2\phi} T_{tt}^{C} \\
(rr) ~& &~ - \frac{1}{2}(\ddot{\sigma} + \delta'' ) - \frac{
(\tilde{r} + r_{s}/2)}{{(\tilde{r} + r_{s})}^{2}} \sigma'
- 2 \gamma'' =
\pi e^{2\phi} T_{rr}^{C} \\
(tr) ~& &~ - \frac{(\tilde{r} + r_{s}/2)}{{(\tilde{r} + r_{s})}^{2}}
\dot{\sigma} - 2{(\dot{\gamma})}' =
\pi e^{2\phi} T_{tr}^{C} \\
(\theta \theta ) ~& &~ \frac{\tilde{r}}{2} ( \delta' + \sigma' )
- \frac{\tilde{r}r_{s}}{{(\tilde{r} + r_{s})}^{2}} \sigma
+ \frac{r_{s}}{{(\tilde{r} + r_{s})}^{2}} \rho = 0
   \end{eqnarray}
where prime and dot denote differentiation with respect to
$r^{*}$ and $t$ respectively.
The 4D energy momentum tensor is that of
the shock wave, and can be simply gotten from the 2D one:
$ ^{(4D)}T_{\mu \nu}^{M} =$$ ^{(2D)}T_{\mu \nu} / 4\pi{(\tilde{r}
+ r_{s} )}^{2} $, where $ 4\pi {( \tilde{r} + r_{s} )}^{2} $ is
the surface factor relating the 2D and 4D densities. Using the
2D results, we find (droping the 4D superscripts)
$ T_{tt}^{C} = T_{rr}^{C} = T_{tr}^{C} = T_{++} $, so
   \begin{equation}
e^{2\phi}T_{tt}^{C} = e^{2\phi}T_{rr}^{C} = e^{2\phi}T_{tr}^{C}
= \frac{x^{+}_{0} \epsilon}{4\pi \tilde{r} (\tilde{r} + r_{s} )}
\delta (x^{+} - x^{+}_{0})
   \end{equation}
where $ x^{\pm} \equiv \pm exp(\pm u^{\pm}) =
\pm exp[\pm (t \pm r^{*})]  $ .
The r.h.s of (26)-(29) is much smaller then 1, which is consistent
with $ \frac{\epsilon}{\tilde{r}_{c}} << 1 $.
We are using here the Eddington-Finkelstein coordinates, $(t,r^{*})$,
for which $ \delta = -\sigma $. So we have from (29)
$ \rho = \tilde{r} \sigma $ and from (26) (or (27) using $T_{tt}
= T_{rr}$)
   \begin{equation}
\frac{1}{2}(\ddot{\sigma} - \sigma'') -
\frac{(\tilde{r} + r_{s}/2)}{{(\tilde{r} + r_{s})}^{2}} \sigma'
- 2 \ddot{\gamma} = \pi e^{2\phi}T_{tt}^{C} \\
   \end{equation}
The two equations (28) and (31) determine the two
functions $ \sigma $ and $ \gamma $. There are two boundary
conditions: the solution must vanish as $\tilde{r} \rightarrow
\infty $, and coincide with the 2D one at $\tilde{r} =
\tilde{r}_{c} $ . The corresponding solution is
   \begin{eqnarray}
\delta = -\sigma & \simeq & \frac{\epsilon r_{s}}{\tilde{r}}
\Theta( u^{+}-u^{+}_{0} ) \\
\gamma & \simeq & 0 \\
\rho & \simeq & 0
   \end{eqnarray}
which is the expected linearised form of the collapsing
star.

\section{Hawking Radiation in a Classical Background}
Consider first the 2D case [4].
Using the trace anomaly one can calculate the Hawking radiation
in the linear dilaton vacuum (the $m=0$ vacuum)
   \begin{equation}
T^{Q}_{++} = 0 ~~,~~
T^{Q}_{--} = \frac{\kappa \lambda^{2}}{4}
\left[ {\left( 1 - \frac{\epsilon}{2\hat{r}} \right)}^{2}
- {\left( 1 - \frac{x^{-}_{H}}{x^{-}} \right)}^{2} \right]
   \end{equation}
where $ x^{-} = x^{-}_{H} $ is the horizon, $ x^{-}_{H} = \frac{
m}{\lambda x^{+}_{0}} $ , and $\kappa \sim \frac{N}{12}$ in
the large $N$ limit
\footnote{$\kappa $ depends on the quantization scheme that one uses
[5-7,15]}.
On $ I_{R}^{+} $ , $ x^{+} \rightarrow \infty $ $ (\hat{r}
\rightarrow \infty $ ) ,
one gets
   \begin{equation}
T^{Q}_{--} \rightarrow \frac{\kappa {\lambda}^{2}}{4} \left(
1 - \left( 1 - \frac{x^{-}_{H}}{x^{-}} \right)^{2} \right)
   \end{equation}
and so for late times ($ x^{-} \rightarrow x^{-}_{H} $), one gets
the thermal Hawking radiation with the temperature $ T_{H} =
\lambda / 2\pi $.

Notice that because $\frac{\epsilon}{\hat{r}_{c}} <<1$, (35)
and (36) are almost the same. This
means that there is almost no redshifting\footnote{ In 2D
there is no surface factor between the
energy and the energy density.} in $ \hat{r} > \hat{r}_{c} $ .

Because at $ \hat{r}=\hat{r}_{c} $
the 2D is a good approximation, the 4D energy flux at
$ \tilde{r}_{c} = \hat{r}_{c}$
will be (35) (multiply by a surface factor). For $\tilde{r} >
\tilde{r}_{c}$, it would seem that one should solve the 4D equations.
But as we are going to see now, it is unnecessary. The 4D radiation
at $\tilde{r} = \tilde{r}_{c}$ will be almost the same as at
infinity, so one can use the 2D radiation.
Consider late time radiation ($x^{-} \rightarrow x^{-}_{H}$).
The 4D radiation at $\tilde{r} \rightarrow \infty$, is related
to the Hawking temperature.
As we know, this temperature goes like $M^{-1}$,
where $M$ is the 4D mass; the 4D Hawking temperature might be
quite different from the
2D one, and if this were the case, the 4D energy momentum tensor at
$ \tilde{r}_{c} $ would be quite different than at infinity.
But this is not the case. The 4D and 2D Hawking
temperature are not very different, and in the $ \epsilon \rightarrow
0 $ ``limit" they are exactly the same.
There are several ways to calculate the Hawking
temperature. One can use the $r-t$ part of the exact 4D solution (5),
and calculate the Hawking radiation in the Israel-Hawking vacuum,
or one can use the surface gravity ${\cal K}$,
   \begin{equation}
T_{H} = \frac{1}{2\pi} {\cal K} =
 \frac{1}{2\pi} \left( lim_{r\rightarrow \infty} g_{tt}^{-1/2}
\right) \left( {(g_{tt}g_{rr})}^{-1/2} \partial g_{tt} / \partial r
\right)(r=r_{H})
   \end{equation}
In the 2D case one get $ T^{(2D)}_{H} = \lambda / 2 \pi $, and in
4D, $ T^{(4D)}_{H} = {( 8 \pi M e^{\phi_{0}} )}^{-1} $. But using
(4) and (15) we see that
   \begin{equation}
T_{H}^{(2D)} = (1+\epsilon){(1-\epsilon)}^{1/2} T_{H}^{(4D)}
   \end{equation}
So (up to $\epsilon$, which we neglect relative to $1$)
the two temperature are the same.
This means that the 4D radiation at $\tilde{r}_{c}$ is almost the
same as at infinity, so (as in the 2D case)
there is almost no redshifting in $\tilde{r} > \tilde{r}_{c} $. This
is consistent with (5), because if $\epsilon / \tilde{r}_{c}
<< 1 $, the 4D metric for $\tilde{r} > \tilde{r}_{c} $ is
almost the ``vacuum metric" (the extremal black hole metric).

All the above considerations assume a classical background metric
(the classical collapsing star), without back-reaction; we next
come to this.

\section{4D and 2D back-reaction}
In the 2D case, one can deal with the back-reaction calculations
much more easily than in the 4D case. But even in this simple 2D
world, there are no known exact solutions to the original CGHS
model [4], though there are some numerical ones [8,9].
But one can extend
the CGHS model and find exact results [5-7]. We will not describe
the details here, but just say that those results describe
formation and evaporation of a black hole, ending with a naked
singularity. An interesting result is that the quantum energy
momentum tensor (describing the Hawking radiation) is exactly
the same as (36), the one that we get using the classical
background.

Another thing is that
the semiclassical approximation is valid as long as $ \frac{m}{
\lambda} > \kappa \sim \epsilon = \frac{m}{\lambda^{3}} $ [16].
This means that $ \lambda^{2} >> 1 $, or equivalently
that the 4D coupling at
infinity is small, $ G^{(4d)}(\tilde{r} \rightarrow \infty )
\simeq \lambda^{-2} << 1 $ . So although we have a very small
4D $f$-shock wave ($\epsilon <<1$), the corresponding 2D mass
$m$, is not small in this weak coupling picture, unlike in
[17], in which a very small 2D shock wave was studied.

The simple S-wave picture is that the 4D metric (for $\tilde{r}
> \tilde{r}_{c}$) should not change much after the Hawking
radiation is started, because if it changes, it is likely that
there will be a big redshift, and the Hawking
radiation at $\tilde{r}_{c}$ will be much different than at
infinity. In that case the only way to calculate it is to
solve the complicated 4D equations (including the back-reaction).
This is of course beyond the scope of this paper, and it
is not clear at all if that 4D solution can be
matched to the 2D one.
Assuming small perturbations of the solution, we can still use the
linearised equations. Assuming S-wave scattering we can use
the perturbations (25), and the linearised form of (26)-(29),
where $T_{\mu \nu}^{C}$ should be replaced by $T_{\mu \nu}^{
M} = T_{\mu \nu}^{C} + T_{\mu \nu}^{Q}$. Using (30) and (36)
we get
   \begin{eqnarray}
e^{2\phi}T_{tt}^{M} &=& e^{2\phi}T_{rr}^{M} =
\frac{1}{4\pi(\tilde{r} + r_{s})} \left[ \frac{x^{+}_{0} \epsilon}{
\tilde{r}} \delta(x^{+}-x^{+}_{0}) +
\frac{\kappa}{4 \tilde{r}} \left( \frac{2x^{-}_{H}}{x^{-}}
- {\left( \frac{x^{-}_{H}}{x^{-}} \right)}^{2} \right) \right]
\nonumber \\
e^{2\phi}T_{tr}^{M} &=&
\frac{1}{4\pi(\tilde{r} + r_{s})} \left[ \frac{x^{+}_{0}\epsilon}{
\tilde{r}}\delta(x^{+}-x^{+}_{0}) -
\frac{\kappa}{4 \tilde{r}} \left( \frac{2x^{-}_{H}}{x^{-}} -
{\left( \frac{x^{-}_{H}}{x^{-}} \right)}^{2} \right) \right]
   \end{eqnarray}
The first part on the r.h.s of (39) is the classical
shock wave part, and the second is the Hawking radiation
part. As we saw in section 3, the shock wave part is much
smaller than one, but what about the second part?
It must also be much smaller than 1. This means that
$\kappa \sim \epsilon
<< 1$ . Very small $\kappa$ means very small Hawking radiation.
In that case the solution of (28) and (31) with (39), is
   \begin{equation}
\delta = - \sigma \simeq \frac{\epsilon r_{s}}{\tilde{r}}
\Theta (u^{+}-u^{+}_{0}) + \frac{\kappa}{4 \tilde{r}}
( x^{-}_{H} e^{-u^{-}} - \frac{1}{4}{(x^{-}_{H})}^{2}
e^{-2u^{-}} + c )
   \end{equation}
where $c$ is a constant to be determined by the continuity
condition at $ \tilde{r} = \tilde{r}_{c} $.
Can (40) be continuously matched to the exact 2D solutions?
Consider the exact solution of [7]. The metric is
$g_{++}=g_{--}=0 ~,~ g_{+-} = - \frac{1}{2}e^{2\rho} $ ,
where $ e^{-2\phi} = e^{-2\rho}$, and
   \begin{eqnarray}
e^{-2\phi} + \frac{\kappa}{2} \phi &=& - \lambda^{2}
x^{+} x^{-} - \frac{\kappa}{4} ln (-\lambda^{2}x^{+}x^{-})
\nonumber \\
 & & - \frac{m}{\lambda x^{+}_{0}} (x^{+} - x^{+}_{0})
\Theta (x^{+} - x^{+}_{0})
   \end{eqnarray}
The metric in the asymptotically flat coordinates is
   \begin{equation}
\tilde{g}_{+-} = -x^{+}(x^{-}-x^{-}_{H}) g_{+-}
   \end{equation}
Using (13), (41) and (42) we get
   \begin{equation}
\tilde{g}_{tt} = \frac{1}{\lambda^{2}} \left( -1 +
\left( \frac{\epsilon}{2 \tilde{r}} + \frac{\kappa}{4
\tilde{r}} ln (- 2 \tilde{r} / x^{+} x^{-}) \right)
\Theta (x^{+}-x^{+}_{0}) \right)
   \end{equation}
So
   \begin{equation}
\delta_{(2D)} =
\left( \frac{\epsilon}{2 \tilde{r}} + \frac{\kappa}{4
\tilde{r}} ln (- 2 \tilde{r} / x^{+} x^{-}) \right)
\Theta (x^{+}-x^{+}_{0})
   \end{equation}
The first term of (44) is exactly the one in (40), but what
about the second term? Using (13), (41) and the fact that
$ \kappa << 1 $ , we get $ 2 \tilde{r} \simeq - x^{+}(
x^{-} - x^{-}_{H}) $, and
   \begin{equation}
\delta_{(2D)} \simeq
\left( \frac{\epsilon}{2 \tilde{r}} + \frac{\kappa}{4
\tilde{r}} ln \left(1 +  \frac{x^{-}_{H}}{x^{-}} \right) \right)
\Theta (x^{+}-x^{+}_{0})
   \end{equation}
Now we see that we can match (45) and (40) only if
$ |x^{-}| >> |x^{-}_{H}| $ (and $ c=0 $).
Only at the beginning of the Hawking radiation can the 2D solution
be matched to the linearised 4D one. When $x^{-}$ approaches
$x^{-}_{H}$, the linearization breaks down, and one can no longer
match the solutions.

In the case of [5,6], the linear dilaton is not a solution to
the semiclassical equations even for $x^{+}<x^{+}_{0}$, so
the matching is more  problematic. But still one can try to
match the solutions at the begininning of the Hawking radiation
process. The solution (of theory I [6]) for $x^{+}>x^{+}_{0}$ is
   \begin{eqnarray}
2 \Omega (\phi) &=& \frac{1}{\kappa} e^{u^{+}}(e^{-u^{-}} - m
/ \lambda ) - \frac{1}{4}(u^{+} - u^{-}) + T + \frac{m}{\lambda
\kappa} + \frac{1}{2} ln(\kappa / 4e) \nonumber \\
\rho &=& - ln (\lambda) - \frac{1}{\kappa}e^{-2\phi} + 2 \Omega
+ \frac{1}{2}(u^{+}-u^{-}) - \frac{1}{2}ln(\kappa / 4e)
   \end{eqnarray}
where
   \begin{equation}
\Omega(\phi) = \frac{e^{-\phi}}{2 \sqrt{\kappa}} \sqrt{
\frac{e^{-2\phi}}{\kappa} - 1 } - \frac{1}{2} ln \left(
\frac{e^{-\phi}}{ \sqrt{\kappa}} + \sqrt{
\frac{e^{-2\phi}}{\kappa} - 1 } \right)
   \end{equation}
For $ \kappa << 1 $, we get from (47) (and $ e^{-2\phi} = 2
\hat{r} $ )
   \begin{equation}
\Omega(\phi) \simeq \frac{e^{-2\phi}}{2 \kappa} + \frac{1}{4}
( ln(\kappa/ 4) - 1 ) + \frac{\kappa}{8 \hat{r}} + \frac{\phi}{2}
   \end{equation}
and from (46) and (48) we get
   \begin{equation}
g_{tt} = - e^{2\rho} = - \frac{e^{(u^{+}-u^{-})}}{
2 \lambda^{2} \hat{r}} \left( 1 + \frac{\kappa}{4 \hat{r}} \right)
   \end{equation}
For very small $ |x^{-}_{H} / x^{-} | $ we get
   \begin{equation}
g_{tt} \simeq \frac{1}{\lambda^{2}} \left( -1 + \frac{\epsilon}{
2 \hat{r}} - \frac{\kappa}{4 \hat{r}} \right)
   \end{equation}
So
   \begin{equation}
\delta_{(2D)} \simeq \frac{\epsilon}{2 \hat{r}} - \frac{\kappa}{
4 \hat{r}}
   \end{equation}
We see that (51) can be matched to (40) if $ c = 1 $,
and only for very very small $ |x^{-}_{H} / x^{-}| $ . The
reason that in this case the matching is to order zero in
$ |x^{-}_{H} / x^{-}| $, while for (45) it was to first
order, is that for $ x^{+} < x^{+}_{0} $  the solution of [6]
is the linear dilaton only to zero order. One can get similar
results for their theory II.

At $\tilde{r} = \tilde{r}_{c}$, and for $x^{+}>x^{+}_{0}$ the
minimum value of $|x^{-}_{H} / x^{-}|$ goes like $ \tilde{r}_{c}
/ \frac{m}{\lambda} $, so the matching is possible only if
$ \tilde{r}_{c} >> \frac{m}{\lambda} $ . This can be consistent
with $ \frac{m}{\lambda} > \kappa $, because $ \tilde{r}_{c}
>> \kappa $ (remember that we keep terms up to first order in
$\kappa / \tilde{r}_{c}$). So in the case of [7], we can match
the solutions (of course only at the beginning of the Hawking
radiation process), but in the case of [5,6] it is again
problematic.

According to our results it
seems that indeed only a small 2D mass is consistent with a
small 4D $f$-shock wave. In that case one should get the results
of [17].

\section{Conclusions}
In this paper we studied the 4D interpretation of the 2D
evaporating black holes. The 4D almost extremal black hole has the
structure of $ \Sigma^{4} = \Sigma^{2} \times S^{2} $ down the
throat, where $\Sigma^{2}$ is the 2D (Witten) black hole.
The 4D collapsing black hole, formed by a shock wave, has the
same product structure (down the throat), where in that case
$\Sigma^{2}$ is the collapsing 2D black hole. The 4D Hawking
radiation in this classical background gives exactly the
2D Hawking radiation in $\Sigma^{2}$ [18].

When back-reaction is taken into account, the picture
changes. A simple 4D S-wave scattering is consistent with the
2D solutions (down the throat) only if the Hawking radiation
is very small\footnote{A small $\kappa$ can
be consistent with a large $N$. If we define the measure of
all the fields [15], with the metric
$\hat{g} = exp(-2\alpha \phi) g $ , then $ \kappa = (1-\alpha)
(N-24)/12 $. In the limit $\alpha \rightarrow 1 $ , $\kappa$
will not vanish only if $N \rightarrow \infty $ (such that
$(1-\alpha ) N \rightarrow const. $ ).}, $\kappa << 1$,
and only at the beginning of the radiation.
Just before an apparent horizon forms ($
x^{-}=x^{-}_{H}$) the linear approximation breaks down, and the
simple 4D  S-wave scattering picture is no longer consistent.
The amount of energy carried by the radiation at that point,
is of order of $\kappa$, which is of the order of $\epsilon$,
the energy carried by the shock wave. So exactly when the
the problem of a positive define 2D mass arises [10-13], the
4D picture breaks down. Perhaps the 4D consideration may
``save us" from the 2D problems. For example the 4D interpretation
is consistent if $\kappa = 0$, and indeed the 2D ($\kappa = 0$)
case is probably consistent [19].

The fact that the 4D picture breaks down even before an
apparent horizon is formed means that the 4D picture could
be quite different than the 2D evaporating picture. It
seems reasonable to believe that the 4D black hole will
radiate the energy of the shock wave and will return to
extremality [16]. Unlike the Reissner-Nordstrom case,
in the dilatonic case, this process will not necesarily
lead to information loss.

If a full 4D back-reaction calculation can be consistent
with a throat region, and if these calculations will be free
from positive energy problems, then the 2D solution must
be consistent as well. The 2D theory that will give this
solution is still (9). But the boundary conditions (at
$\hat{r}=\hat{r}_{c}$) will be different. It could be
interesting to find those boundary conditions (without
solving the 4D theory), and to see if it corresponds to
a reasonable 4D picture.

\vspace{1cm}
{\bf Acknowledgments}\\
I would like to thank Stanley Deser and Alan Steif for very
helpful discussions.

\vspace{1.5cm}
{\bf References}
\begin{enumerate}
\item D.Garfinkle, G.Horowitz and A.Strominger, {\em Phys.Rev.Lett.}
{\bf 67}, 3140 (1991) \\
\hspace{0.5cm} G.W.Gibbons and K.Maeda, {\em Nucl.Phys.}{\bf B298},
741 (1988)
\item S.B.Giddings and A.Strominger, {\em Phys.Rev.}{\bf D46}
627 (1992)
\item E.Witten, {\em Phys.Rev.}{\bf D44}, 314 (1991) \\
\hspace{0.5cm} S.Elizur, A.Forge and E.Rabinovici,
{\em Nucl.Phys.}{\bf B359}, 581 (1991) \\
\hspace{0.5cm} G.Mandal, A.Sengupta and S.Wadia, {\em Mod.Phys.Lett}
{\bf A6}, 1685 (1991)
\item C.G.Callan,S.B.Giddings,J.A.Harvey and A.Strominger,
{\em Phys.Rev.}{\bf D45}, R1005 (1992)
\item S.P.de Alwis, University of Colorado Report No. COLO-HEP-280,
hepth@xxx/920569 (1992); report No. COLO-HEP-284, hepth@xxx/ 9206020
(1992)
\item A.Bilal and C.G.Callan, Princeton University Report No.
PUPT-1320, hepth@xxx/9205089 (1992)
\item J.G.Russo, L.Susskind and L.Thorlacius, {\em Phys.Rev.}
{\bf D46}, 3444 (1992)
\item S.W.Hawking, {\em Phys.Rev.Lett.}{\bf 69}, 406 (1992)
\item S.W.Hawking and J.M.Stewart, Cambridge University Report
(1992), DAMTP/R-92/37
\item S.B.Giddings and A.Strominger, Santa Barbara Report No.
UCSBTH-92-28, hepth@xxx/9207034 (1992)
\item Y.Park and A.Strominger, Santa Barbara Report No.
UCSBTH-92-39, hephth@xxx/9210017
\item S.P. de Alwis, Colorado University Report No. COLO-HEP-309,
hepth @xxx/9302144 (1993)
\item A.Bilal and I.I.Kogan, Princeton University Report No.
PUPT-1379, hepth@xxx/9303119 (1993)
\item S.W.Hawking, {\em Comm.Math.Phys.}{\bf 43}, 199 (1975)
\item A.Strominger, Santa Barbara Report No. UCSBTH-92-81,
hepth@xxx/ 9205028 (1992) to be published in {\em Phys.Rev.}{\bf D}
\item L.Susskind and L.Thorlacius, Stanford University Report No.
SU-ITP-92-12, (1992) to be published in {\em Nucl.Phys.}{\bf B}
\item A.Strominger and S.P.Trivedi, NSFITP-93-15 CALT-68-1851,
hepth@ xxx/9302080 (1993)
\item S.B.Giddings and W.M.Nelson, {\em Phys.Rev.}{\bf D46},
2486 (1992)
\item E.Verlinde and H.Verlinde, PUPT-1380, IASSNS-HEP-93/8

\end{enumerate}

\end{document}